\documentclass[aps,pra,showpacs,twocolumn,amsmath,amssymb,superscriptaddress,nofootinbib]{revtex4}
\usepackage{graphicx}
\usepackage{bm}
\usepackage{amssymb}
\usepackage{amsmath}
\usepackage{latexsym}
\usepackage{color}
\newcommand{\vc}[1]{\mbox{\boldmath $#1$}} 
\newcommand{\ind}[1]{_{#1}}    
\newcommand{\indrm}[1]{_{\mathrm {#1}}}    
\newcommand{\dirate}{{\mathcal D}}   
\newcommand{\sgn}{s}   

\newcommand{\fspace}{P}   
\newcommand{\thinlens}{L}   
\newcommand{\crystal}{C}   
\newcommand{\crystalsp}{K}

\newcommand{\focus}{F}   
\newcommand{\dcomm}[1]{b_{\cup_{\ind{#1}}}}   
\newcommand{\acomm}[1]{1/b_{\cup_{\ind{#1}}}}
\newcommand{\bcomm}[1]{B_{\cup_{\ind{#1}}}}      
\newcommand{\gcomm}[1]{G_{\cup_{\ind{#1}}}}      
\newcommand{\fcomm}[1]{\dirate_{\cup_{\ind{#1}}}}
\newcommand{\elmt}{e}   
\newcommand{\cmb}{C}   
\newcommand{\an}{R}   
\newcommand{\mo}{D}   
\newcommand{\etra}{\varepsilon}
\newcommand{\deis}[1]{\Delta E_{\indrm{#1}}^{({\mathrm s})}}   
\newcommand{\dais}[1]{\Delta \theta_{\indrm{#1}}^{({\mathrm s})}}   
\newcommand{\dei}[1]{\Delta E_{\indrm{#1}}}   
\newcommand{\dai}[1]{\Delta \theta_{\indrm{#1}}}   
\begin{document}  
\title{X-ray echo spectroscopy}
\author{Yuri Shvyd'ko}
\email{shvydko@aps.anl.gov}
\affiliation{Advanced Photon Source, Argonne National Laboratory,   Argonne, Illinois 60439, USA} 

\begin{abstract} 
  X-ray echo spectroscopy, a counterpart of neutron spin-echo, is
  being introduced here to overcome limitations in spectral resolution
  and weak signals of the traditional inelastic x-ray scattering (IXS)
  probes.  An image of a point-like x-ray source is defocused by a
  dispersing system comprised of asymmetrically cut specially arranged
  Bragg diffracting crystals.  The defocused image is refocused into a
  point (echo) in a time-reversal dispersing system.  If the defocused
  beam is inelastically scattered from a sample, the echo signal
  acquires a spatial distribution, which is a map of the inelastic
  scattering spectrum. The spectral resolution of the echo
  spectroscopy does not rely on the monochromaticity of the x-rays,
  ensuring strong signals along with a very high spectral resolution.
  Particular schemes of x-ray echo spectrometers for 0.1--0.02-meV
  ultra-high-resolution IXS applications (resolving power $> 10^8$)
  with broadband $\simeq$~5--13~meV dispersing systems are introduced
  featuring more than $10^3$ signal enhancement. The technique is
  general, applicable in different photon frequency domains.
\end{abstract}



\maketitle

\section{Introduction}

The spectroscopic signal strength decreases abruptly with improving
spectral resolution.  An x-ray echo spectroscopy, introduced here,
offers a potential for achieving much higher, yet unattainable,
spectral resolution in the hard x-ray regime without compromising the
signal strength.

The origin of the proposed technique is in spin echo, a phenomenon
discovered by Erwin Hahn in 1950 \cite{Hahn50}. Spin echo is the
refocusing in the time domain of the defocused spin magnetization by
time reversal. The spin echo technique is widely used in NMR. There is
a photon echo analog in the optics. Neutron spin echo spectroscopy is
an inelastic neutron scattering technique invented by Mezei, which
uses time reversal of the neutron spin evolution to measure energy
loss in an inelastic neutron scattering process \cite{Mezei80}.
Photon polarization precession spectroscopy was proposed recently by
R\"ohlsberger \cite{Roehle14} for studies of spin waves that exhibit
similarities to the neutron spin echo.  Fung et~al. proposed a
space-domain analog of the echo spectroscopy for resonant inelastic
soft x-ray scattering applications \cite{Fung04}.  Defocusing and
refocusing of the spectral components is achieved by angular
dispersion from curved diffraction gratings.  This approach has been
recently demonstrated by Lai et al. \cite{Lai14}.

Here, we propose a hard x-ray version of the echo spectroscopy, which
can be applied for non-resonant and resonant high-resolution inelastic
x-ray scattering applications.  Diffraction gratings are not practical
in the hard x-ray regime. However, as was demonstrated in
\cite{Shvydko-SB,SLK06}, the angular dispersion in the hard x-ray
regime can be achieved by Bragg diffraction from asymmetrically cut
crystals or from special arrangements of asymmetrically cut crystals
\cite{SSM13,Shvydko15}, which are a hard x-ray analog of the optical
diffraction gratings and optical prisms. In the space-domain
echo-spectrometer proposed here, an image of a point-like x-ray
source is defocused by a dispersing system comprised of asymmetrically
cut Bragg diffracting crystals. The defocused image is refocused into
a point (echo) in a time-reversal dispersing system. We show, if the
defocused beam is inelastically scattered from a sample, the echo
signal acquires a spatial distribution, which is a map of the energy
transfer spectrum in the scattering process.  The spectral resolution
of the echo spectroscopy does not rely on the monochromaticity of the
x-rays, thus ensuring strong signals along with a very high spectral
resolution.

In the present paper, we use an analytical ray-transfer matrix
approach and the dynamical theory of x-ray diffraction in crystals to
calculate and analyze the performance of a generic echo spectrometer
comprised of defocusing and refocusing dispersing elements, to derive
conditions for refocusing and expressions for the spectral resolution
of the echo spectrometer. Specific designs of the hard x-ray echo
spectrometers are introduced with a spectral resolution $\Delta
\etra=0.1-0.02$~meV at photon energies $E\simeq 9.1$~keV and
$E\simeq 4.6$~keV, comprised of defocusing and refocusing systems with
multi-crystal inline dispersing elements featuring both large
cumulative dispersion rates $\fcomm{} \gtrsim 25-60~\mu$rad/meV,
transmission bandwidths $\Delta E_{\ind{\cup}}\simeq 5-13$~meV, and a
dynamical range $\Delta E_{\ind{\cup}}/\Delta \etra\simeq
100-500$.  
 Because of much greater dispersion rates which are feasible
 in the crystal systems, as compared to the diffraction gratings, the
 spectral resolving power in the hard x-ray regime can be as large as
 $\gtrsim 10^8-10^9$, i.e., more than three orders of magnitude higher
 than in the soft x-ray regime.

\section{Theory}

\begin{figure}[t!]
\includegraphics[width=0.5\textwidth]{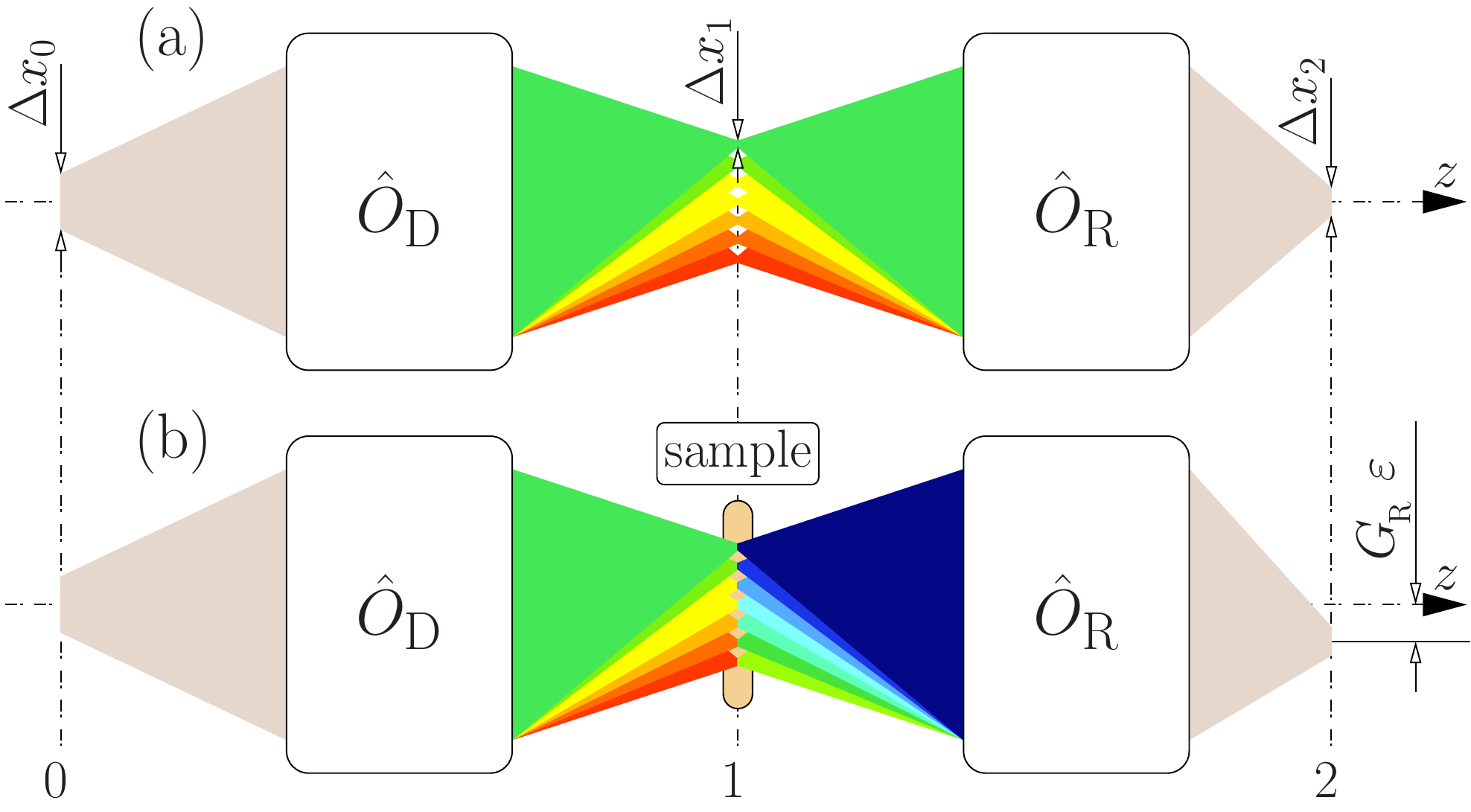}
\caption{Graphical presentation of the echo spectroscopy
  principles. (a) Photons from a source with a linear size $\Delta
  x_{\indrm{0}}$ in the reference source plane $0$ are focused onto a
  spot $\Delta x_{\indrm{1}}$ in the intermediate image plane $1$ by a
  focusing-dispersing system $\hat{O}_{\indrm{\mo}}$. Each spectral
  component, indicated by different color, is
  focused at a different location due to dispersion in
  $\hat{O}_{\indrm{\mo}}$. All spectral components of the x-rays are
  refocused by a consecutive time-reversal focusing-dispersing system
  $\hat{O}_{\indrm{\an}}$ onto the same spot $\Delta x_{\indrm{2}}$
  (echo) in the image plane $2$. (b) Inelastic x-ray scattering with
  an energy transfer $\etra$ (indicated by changed color) 
  from a sample in the reference plane $1$ results in
  a lateral shift $G_{\indrm{\an}} \etra$ of the echo signal
  equal for all spectral components.
  Here  $G_{\indrm{\an}}$ is a linear dispersion rate in the time-reversal
  system $\hat{O}_{\indrm{\an}}$. Spectral resolution of the x-ray
  echo spectrometer is $\Delta \etra =\Delta
  x_{\indrm{2}}/G_{\indrm{\an}}$.  
}
\label{fig001}
\end{figure}

We start here by considering  optical systems featuring a combination
of focusing and energy dispersing capabilities. We assume that such
systems can, first, focus {\em monochromatic} x-rays from a source of
a linear size $\Delta x_{\ind{0}}$ in a source plane (reference plane
$0$ perpendicular to the optical axis $z$ in Fig.~\ref{fig001}) onto
an intermediate image plane (reference plane $1$ in Fig.~\ref{fig001})
with an image linear size $\Delta x_{\ind{1}}= |A| \Delta
x_{\ind{0}}$, where $A$ is a magnification factor of the optical
system. In addition, the system can disperse photons in such a way
that the location of the source image for photons with an energy
$E+\delta E$ is displaced in the image plane by $G \delta E$ from the
location of the image for photons with energy $E$. Here, $G$ is a
linear dispersion rate of the system.  As a result, although
monochromatic x-rays are focused, the whole spectrum of x-rays is
defocused, due to linear dispersion.

We will use the ray-transfer matrix technique
\cite{KL66,MK80-1,Siegman} to propagate paraxial x-rays through such
optical systems and to determine linear and angular sizes of the x-ray
beam along the optical axis. A paraxial ray in any reference plane is
characterized by its distance $x$ from the optical axis, by its angle
$\xi$ with respect to that axis, and the deviation $\delta E$ of the
photon energy from a nominal value $E$.  The ray vector
$\vc{r}_{\ind{0}}=(x_{\ind{0}},\xi_{\ind{0}},\delta E)$ at an input
source plane is transformed to
$\vc{r}_{\ind{1}}=(x_{\ind{1}},\xi_{\ind{1}},\delta
E)=\hat{O}\vc{r}_{\ind{0}}$ at the output reference plane (image
plane), where $\hat{O}=\{ABG;CDF;001\}$ is a ray-transfer matrix of an
optical element placed between the planes  
\footnote{The beam size
  $\Delta x$, the angular spread $\Delta \xi$, and the energy spread
  $\Delta E$ are obtained by the propagation of second-order statistical
  moments, using transport matrices derived from the ray-transfer
  matrices, and assuming zero cross-correlations (i.e., zero mixed
  second-order moments).}.  
Only elastic processes in the optical
systems are taken into account, that is reflected by zero and unity
terms in the lowest row of the ray-transfer matrices.

Focusing of the monochromatic spectral components requires that matrix
element $B=0$. The ray-transfer matrix of any focusing-dispersing
system in a general case therefore reads as
\begin{equation}
\hat{O}  = 
 \{A \,0\,G;\,CDF;001\}
\label{matrix}
\end{equation}
with $A$ and $G$ elements defined above. The system blurs the
polychromatic source image, because of linear dispersion, as mentioned
earlier and graphically presented in Fig.~\ref{fig001}(a).  However,
another focusing-dispersing system
can be used to
refocus the source onto reference plane 2.  Indeed, propagation of
x-rays through the defocusing system $\hat{O}_{\indrm{\mo}}$ and a
second system, which we will refer to as a refocusing or time-reversal
system $\hat{O}_{\indrm{\an}}$ (see Fig.~\ref{fig001}) is given by a
combined ray-transfer matrix
\begin{multline}
\hat{O}_{\indrm{\cmb}}  = \hat{O}_{\indrm{\an}} \hat{O}_{\indrm{\mo}} = \{A_{\indrm{\cmb}}\,0\,G_{\indrm{\cmb}};C_{\indrm{\cmb}}D_{\indrm{\cmb}}F_{\indrm{\cmb}};001\}\\  = \left(\!\! \begin{array}{ccc} A_{\indrm{\an}} A_{\indrm{\mo}}  & 0 &  A_{\indrm{\an}}G_{\indrm{\mo}}+G_{\indrm{\an}}  \\ C_{\indrm{\an}}A_{\indrm{\mo}}\!+\!D_{\indrm{\an}}C_{\indrm{\an}}   & D_{\indrm{\an}} D_{\indrm{\mo}}  & C_{\indrm{\an}}G_{\indrm{\mo}}\!+\!D_{\indrm{\an}}F_{\indrm{\mo}}\!+\!F_{\indrm{\an}} \\ 0 & 0 & 1 \end{array}\!\! \right),
\label{comb}
\end{multline}
and by a ray vector $\vc{r}_{\ind{2}}=(x_{\ind{2}},\xi_{\ind{2}},\delta
E)=\hat{O}_{\indrm{\cmb}}\vc{r}_{\ind{0}}$. 

Here we arrive at a crucial point. If 
\begin{equation}
G_{\indrm{\cmb}}=A_{\indrm{\an}}G_{\indrm{\mo}}+G_{\indrm{\an}}=0,
\label{refocus}
\end{equation}
the linear dispersion at the exit of the combined system vanishes,
because dispersion in the defocusing system is compensated (time
reversed) by dispersion in the refocusing system. As a result, the
combined system refocuses all photons independent of the photon energy
to the same location, $x_{\ind{2}}$ in image plane $2$, to a spot with a
linear size
\begin{equation}
\Delta x_{\ind{2}}= |A_{\indrm{\an}} A_{\indrm{\mo}}| \Delta x_{\ind{0}}\equiv  |A_{\indrm{\an}} | \Delta x_{\ind{1}}, 
\label{magnification}
\end{equation}
as shown schematically in Fig.~\ref{fig001}(a).  Such behavior is an
analog of the echo phenomena. Here, however, it takes place in space,
rather than in the time domain.

Now, what happens if a sample is placed into the intermediate image
plane $1$, [Fig.~\ref{fig001}(b)], which can scatter photons
inelastically? 
In an inelastic scattering process, a photon with an
arbitrary energy $E+\delta E$, changes its value to $E+\delta E+
\etra$. Here $\etra$ is an 
energy transfer
in the inelastic scattering process.  The ray vector
$\vc{r}_{\ind{1}}=(x_{\ind{1}},\xi_{\ind{1}},\delta E)$ before
scattering transforms to
$\vc{r}_{\ind{1}}^{\prime}=(x_{\ind{1}},\xi_{\ind{1}}^{\prime},\delta
E+\etra)$ after inelastic scattering. Propagation of
$\vc{r}_{\ind{1}}^{\prime}$ through the time-reversal system results
in a ray vector
$\vc{r}_{\ind{2}}^{\prime}=(x_{\ind{2}}^{\prime},\xi_{\ind{2}}^{\prime},\delta
E+\etra)= \hat{O}_{\indrm{\an}}\vc{r}_{\ind{1}}^{\prime}$.
Assuming that refocusing condition \eqref{refocus} holds, we come to
a decisive point: all photons independent of the incident photon
energy $E+\delta E$ are refocused to the same location
\begin{equation}
x_{\ind{2}}^{\prime}=x_{\ind{2}}+G_{\indrm{\an}} \etra,\hspace{0.5cm}  x_{\ind{2}}=A_{\indrm{\an}} A_{\indrm{\mo}} x_{\ind{o}},
\label{magnification2}
\end{equation}
which is, however, shifted from $x_{\ind{2}}$ by $G_{\indrm{\an}}
\etra$, a value proportional to the energy transfer $\etra$ in the
inelastic scattering process. The essential point is that, the
combined defocusing-refocusing system maps the inelastic scattering
spectrum onto image plane $2$. The image is independent of the
spectral composition $E+\delta E$ of the photons in the incident beam.

The spectral resolution $\Delta \etra$ of the echo spectrometer is
calculated from the condition, that the shift due to inelastic
scattering $x_{\ind{2}}^{\prime}-x_{\ind{2}}=G_{\indrm{\an}} 
\etra$ is at least as large as the linear size $\Delta
x_{\ind{2}}$ of the echo signal  \eqref{magnification}:
\begin{equation}
\Delta \etra =    \frac{\Delta x_{\ind{2}}}{|G_{\indrm{\an}}|}  \equiv  \frac{|A_{\indrm{\an}}|\Delta x_{\ind{1}} }{|G_{\indrm{\an}}|} \equiv  \frac{|A_{\indrm{\an}} A_{\indrm{\mo}}| \Delta x_{\ind{o}}}{|G_{\indrm{\an}}|}.
\label{resolution}
\end{equation}
Here it is assumed that the spatial resolution of
the detector is better than $\Delta x_{\ind{2}}$.

These results constitute the underlying principle of x-ray echo
spectroscopy. Noteworthy, angular dispersion always results in an
inclined intensity front, i.e., in dispersion both perpendicular to
and along the beam propagation direction \cite{SL12}.  Therefore,
x-rays are defocused and refocused also in the time domain, as in
spin-echo. As a result, inelastic scattering spectra can be also
mapped by measuring time distributions in the detector, given a
short-pulse source.

Perfect refocusing takes place if the linear dispersion of the
combined system
$G_{\indrm{\cmb}}=A_{\indrm{\an}}G_{\indrm{\mo}}+G_{\indrm{\an}}$
vanishes, as in Eq.~\eqref{refocus}. Refocusing can still take place
with good accuracy if $|G_{\indrm{\cmb}}|$ is sufficiently small:
\begin{equation}
|G_{\indrm{\cmb}}| \Delta E_{\cup}\ll \Delta x_{\ind{2}}.
\label{tolerances} 
\end{equation}
Here $\Delta E_{\cup}$ is the bandwidth of x-rays in image plane
$2$. Tolerances on the echo spectrometer parameters and on the sample
shape can be calculated with Eq.~\eqref{tolerances}.

The above approach is general and applicable to any frequency
domain. A particular version was proposed in the soft x-ray domain,
for applications in resonant IXS spectroscopy at $L$-edges of
$3d$ elements \cite{Fung04,Lai14}.  The dispersing elements in the
soft x-ray and visible light domains are diffraction gratings.
with diffraction gratings as dispersing elements \cite{Fung04,Lai14}.

\section{Optical Design}

Diffraction gratings are not practical in the hard x-ray regime.
Extension into the hard x-ray regime is therefore nontrivial.  In this
regard, as was demonstrated in \cite{Shvydko-SB,SLK06}, the angular
dispersion in the hard x-ray regime can be achieved by Bragg
diffraction from asymmetrically cut crystals, i.e., from crystals with
the reflecting atomic planes not parallel to the entrance
surface. This is a hard x-ray analog of the optical diffraction
gratings or optical prisms.  A large dispersion rate is a key for
achieving high spectral resolution in angular-dispersive x-ray
spectrometers \cite{SSS14,Shvydko15}, including echo spectrometers,
see Eq. (6). This is achieved, first, by using strongly asymmetric
Bragg reflections close to backscattering \cite{Shvydko-SB,SLK06},
and, second, by enhancing the single-reflection dispersion rate
considerably by subsequent asymmetric Bragg reflections from crystals
in special arrangements \cite{SSM13} exemplified below.  In the
following two steps, we will show how the principle scheme of a
generic echo spectrometer presented above, can be realized in the hard
x-ray regime by using multi-crystal arrangements as dispersing
elements.

\begin{figure}[t!]
\includegraphics[width=0.5\textwidth]{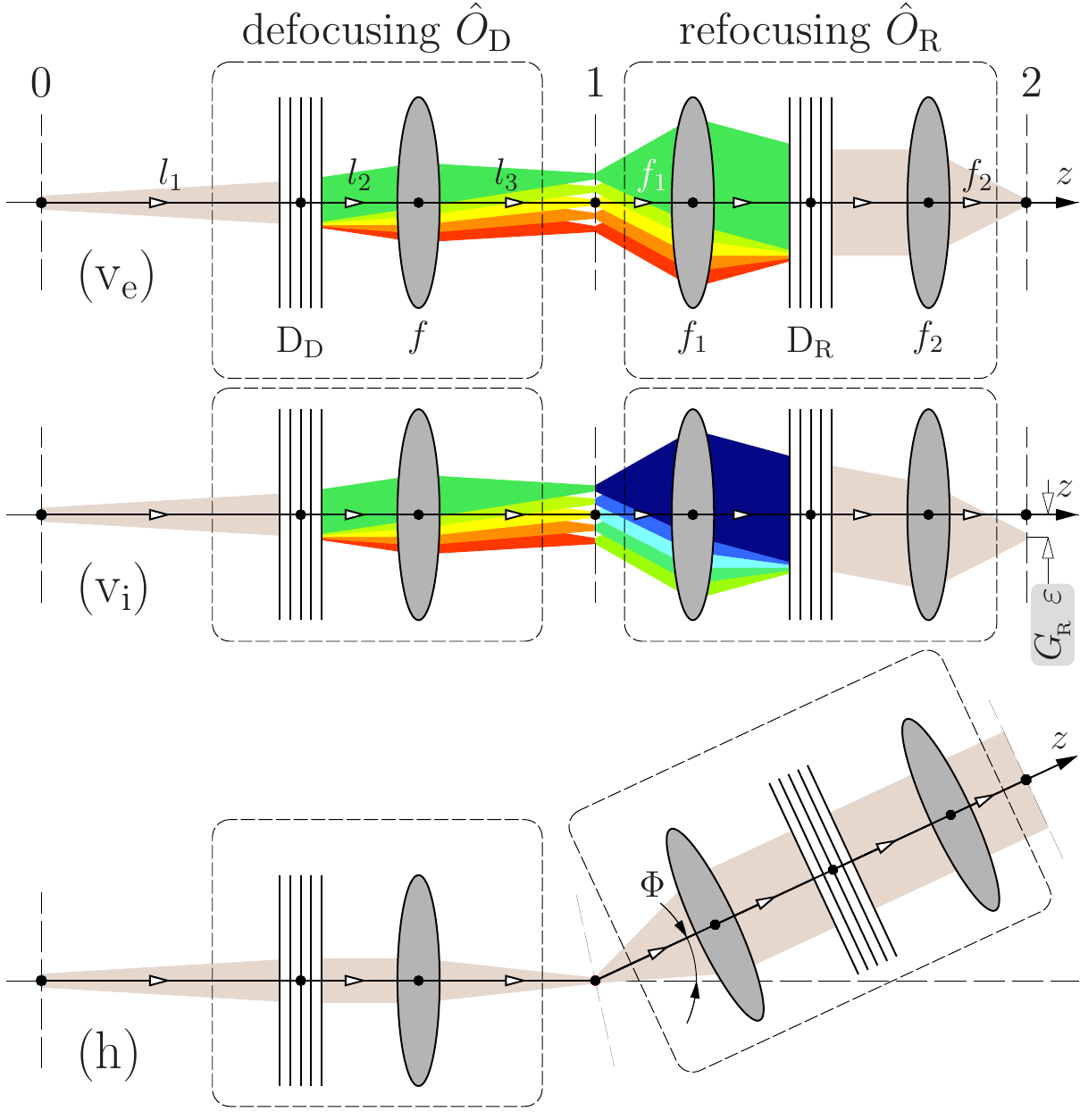}
\caption{Principle optical scheme of an x-ray echo spectrometer,
  comprised of the defocusing $\hat{O}_{\indrm{\mo}}$ and refocusing
  $\hat{O}_{\indrm{\an}}$ dispersing systems; the x-ray source in
  reference plane $0$; the sample in plane $1$; and the
  position-sensitive detector in $2$.  The spectrometer is shown in
  the vertical dispersion plane for elastic (v$_{\indrm{e}}$) and
  inelastic (v$_{\indrm{i}}$) scattering, and in the horizontal
  scattering plane (h) with the refocusing system at a scattering
  angle $\Phi$.  The defocusing system $\hat{O}_{\indrm{\mo}}$
  consists of a dispersing Bragg diffracting (multi)crystal element
  D$_{\indrm{\mo}}$, and of a focusing element $f$.  The refocusing
  system $\hat{O}_{\indrm{\an}}$ is of a spectrograph type comprising
  collimating element $f_{\ind{1}}$; a dispersing Bragg diffracting
  (multi)crystal element D$_{\indrm{\an}}$; and a focusing element
  $f_{\ind{2}}$.  }
\label{fig002}
\end{figure}

In the first step, we propose a principle optical scheme of a hard
x-ray echo spectrometer (Fig.~\ref{fig002}) comprised of the
defocusing $\hat{O}_{\indrm{\mo}}$ and refocusing
$\hat{O}_{\indrm{\an}}$ dispersing systems. The x-ray source is in
reference plane $0$, the sample is in plane $1$, and the
position-sensitive detector is in plane $2$.  The defocusing system
$\hat{O}_{\indrm{\mo}}$ is proposed here as a combination of a Bragg
(multi)crystal dispersing element D$_{\indrm{\mo}}$ and a focusing
element $f$.  As has been shown in \cite{Shvydko15}, such a system can
be represented by a ray-transfer matrix \eqref{matrix} with the
magnification $A_{\indrm{\mo}}$ and linear dispersion
$G_{\indrm{\mo}}$ matrix elements given by
\begin{equation}
A_{\indrm{\mo}}\!=\! - \frac{1}{\dcomm{\mo}}\!\frac{l_{\ind{3}}}{l_{\ind{12}}},\hspace{0.20cm}
G_{\indrm{\mo}}\!=\!\fcomm{\mo}  \frac{l_{\ind{3}} l_{\ind{1}}}{\dcomm{\mo}^2 l_{\ind{12}}},\hspace{0.20cm} l_{\ind{12}}\! =\!\frac{l_{\ind{1}}}{\dcomm{\mo}^2} + l_{\ind{2}}. 
\label{defocusing}
\end{equation}
Here, $l_{\ind{1}}$, $l_{\ind{2}}$, and $l_{\ind{3}}$, are the
distances between the x-ray source, the dispersing element
D$_{\indrm{\mo}}$, the focusing element $f$ with  focal length
$f=(l_{\ind{12}}^{-1}+l_{\ind{3}}^{-1})^{-1}$, and the sample in the
image plane $1$, respectively (Fig.~\ref{fig002}). The dispersing
(multi)crystal system D$_{\indrm{\mo}}$ is characterized by the
cumulative angular dispersion rate $\fcomm{\mo}$, and  cumulative
asymmetry factor $\dcomm{\mo}$, which are defined in \cite{Shvydko15}
(see also Appendix~\ref{raytransfer} and Table~\ref{tab2}).

\begin{figure}
\includegraphics[width=0.5\textwidth]{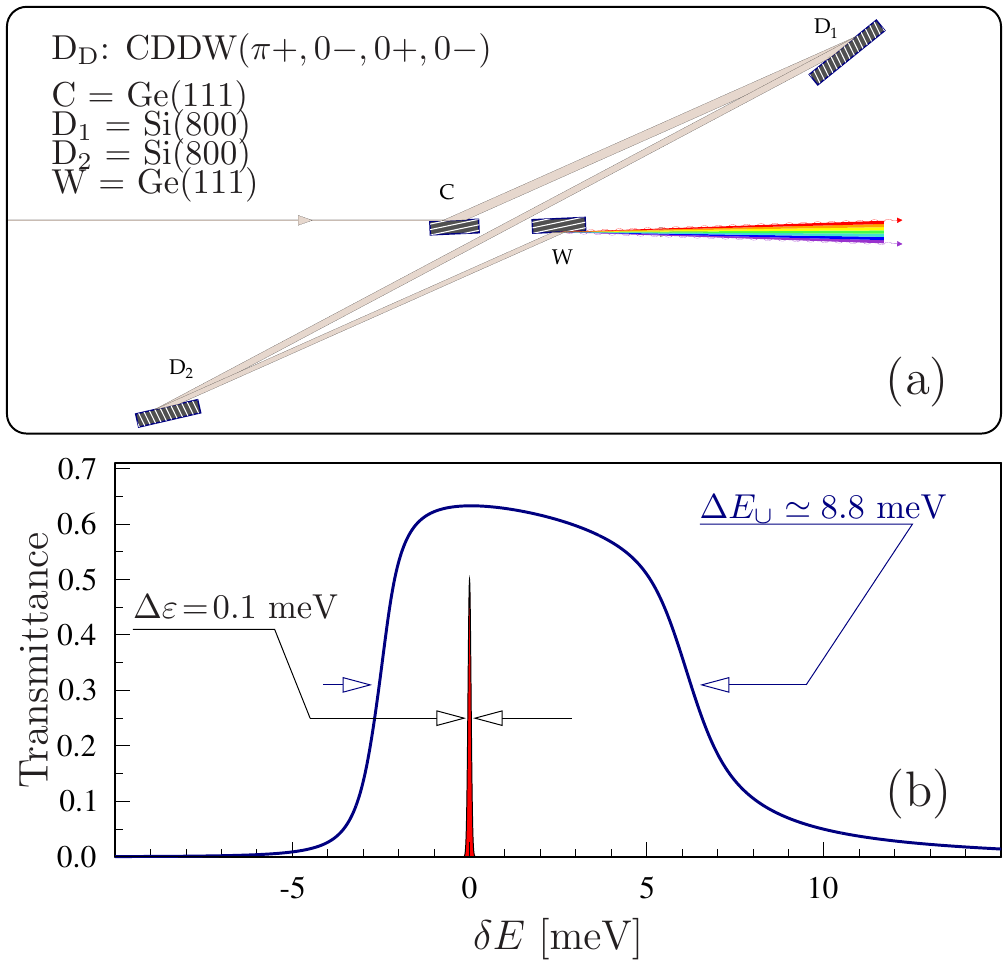}
\caption{Dispersing element D$_{\indrm{\mo}}$ (a) of the defocusing
  system $\hat{O}_{\indrm{\mo}}$ (see Fig.~\ref{fig002}) and its
  spectral transmittance function (b).  D$_{\indrm{\mo}}$ is an
  example of an in-line four-crystal CDDW-type dispersing optic,
  comprised of collimating (C), dispersing (D$_{\ind{1}}$,
  D$_{\ind{2}}$), and wavelength-selecting (W) crystals in a
  ($\pi+$,$0-$,$0+$,$0-$) scattering configuration. With the crystal
  parameters provided in 
  Table~\ref{tab1} of Appendix~\ref{cddwoptic}, 
  the dispersing element D$_{\indrm{D}}$ features a spectral transmission
  function with a $\Delta E_{\ind{\cup}}=8.8$~meV bandwidth (b), a
  cumulative angular dispersion rate $\fcomm{\mo} = -25 ~\mu$rad/meV,
  and a cumulative asymmetry factor $\dcomm{\mo} = 1.4$.  The sharp
  line in (b) presents the 0.1-meV spectral resolution of an
  x-ray echo spectrometer, designed for use with 9.1-keV photons. }
\label{fig003}
\end{figure}

\begin{figure}
\includegraphics[width=0.5\textwidth]{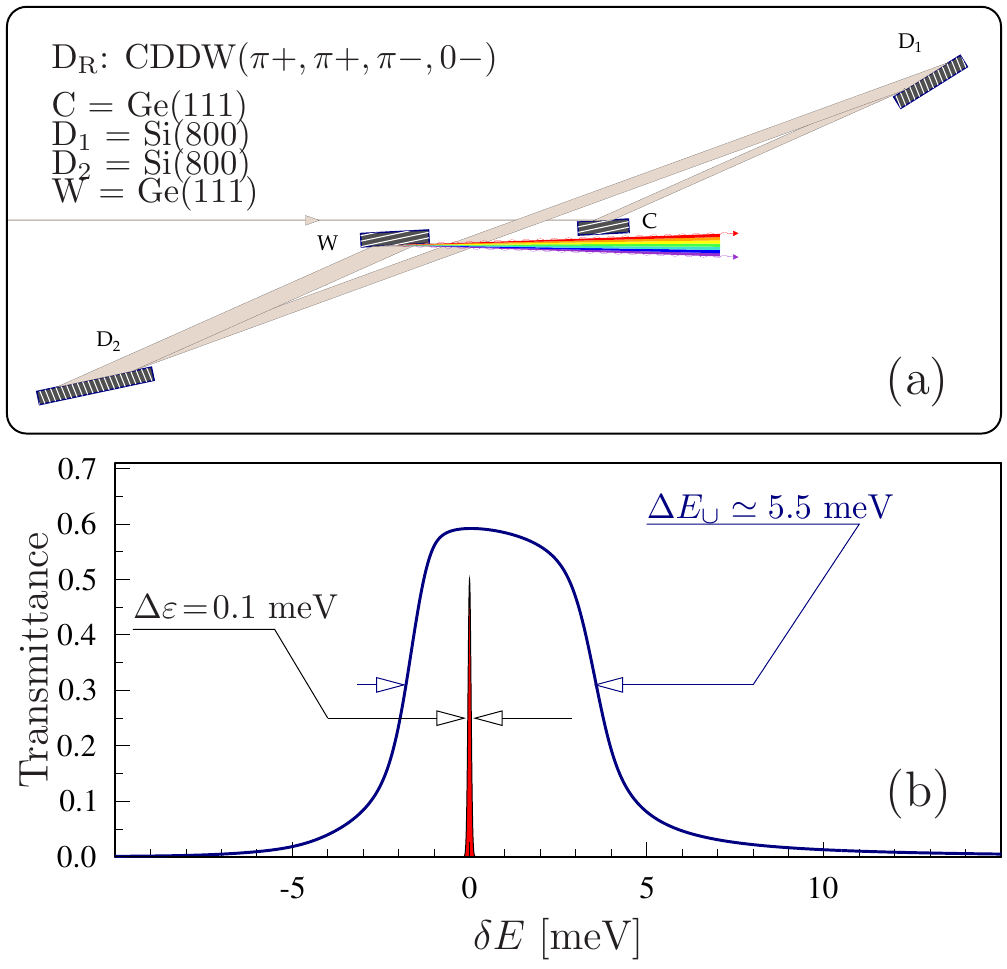}
\caption{Dispersing element D$_{\indrm{\an}}$ (a) of the refocusing
  system $\hat{O}_{\indrm{\an}}$ (see Fig.~\ref{fig002}) and its
  spectral transmittance function (b). Similar to D$_{\indrm{\mo}}$ in
  Fig.~\ref{fig003}, D$_{\indrm{\an}}$ is an example of the in-line
  four-crystal CDDW-type dispersing optic, but, in a
  ($\pi+$,$\pi+$,$\pi-$,$0-$) scattering configuration. With the
  crystal parameters provided in  Table~\ref{tab1} of Appendix~\ref{cddwoptic}, 
  the dispersing element D$_{\indrm{D}}$ features a spectral
  transmission function with a $\Delta E_{\ind{\cup}}=5.5$~meV
  bandwidth (b), a cumulative angular dispersion rate $\fcomm{\mo} =
  -43 ~\mu$rad/meV, a cumulative asymmetry factor $\dcomm{\mo} =
  0.36$, and $\fcomm{\mo}/\dcomm{\mo} = -120 ~\mu$rad/meV. }
\label{fig004}
\end{figure}

For the spectrometer to feature a large throughput, the refocusing
system $\hat{O}_{\indrm{\an}}$ has to be capable of collecting x-ray
photons in a large solid angle scattered from the sample. For this
purpose, we propose using a hard x-ray focusing-dispersing system of a
spectrograph-type considered in \cite{Shvydko15}, and schematically
shown in Fig~\ref{fig002}.  A collimating focusing element
$f_{\ind{1}}$ collects photons in a large solid angle and makes x-ray
beams of each spectral component parallel. The collimated beams
impinge upon the Bragg (multi)crystal dispersing element
D$_{\indrm{\an}}$ with the cumulative angular dispersion rate
$\fcomm{\an}$, and the cumulative asymmetry factor $\dcomm{\an}$. The
focusing element $f_{\ind{2}}$ focuses x-rays in the vertical
dispersion plane onto the detector placed in the image plane $2$.  As
shown in \cite{Shvydko15} 
(see also Appendix~\ref{raytransfer} and Table~\ref{tab2}) such a
system is described by a ray-transfer matrix \eqref{matrix} with the
magnification $A_{\indrm{\an}}$ and linear dispersion
$G_{\indrm{\an}}$ matrix elements given by
\begin{equation}
A_{\indrm{\an}}=-\frac{\dcomm{\an}f_{\ind{2}}\!}{f_{\ind{1}}\!},\hspace{0.5cm}
G_{\indrm{\an}}=\fcomm{\an} f_{\ind{2}} .
\label{refocusing}
\end{equation}
Using Eqs.~\eqref{refocus}, \eqref{defocusing}, and 
\eqref{refocusing} we obtain for the refocusing condition in the hard x-ray echo spectrometer
\begin{equation}
 \frac{l_{\ind{3}} l_{\ind{1}}}{l_{\ind{1}} + \dcomm{\mo}^2 l_{\ind{2}}} \fcomm{\mo} = f_{\ind{1}} \frac{\fcomm{\an}}{\dcomm{\an}}. 
\label{refocus2}
\end{equation}
The dispersing element D$_{\indrm{\mo}}$ can be placed from the source
at a large distance $l_{\ind{1}} \gg \dcomm{\mo}^2 l_{\ind{2}}$. In
this case, the refocusing condition \eqref{refocus2} reads
\begin{equation}
 l_{\ind{3}}\fcomm{\mo} \simeq  f_{\ind{1}} \frac{\fcomm{\an}}{\dcomm{\an}}. 
\label{refocus23}
\end{equation}
For the spectral resolution $\Delta \etra$ of a hard x-ray echo
spectrometer we obtain from Eqs.~\eqref{resolution},
\eqref{defocusing}, and \eqref{refocusing}:
\begin{equation}
\Delta \etra =   \frac{|\dcomm{\an}|}{|\fcomm{\an}|} \frac{\Delta x_{\ind{1}}}{f_{\ind{1}}}. 
\label{resolution2}
\end{equation}
As follows from Eq.~\eqref{resolution2}, the spectral resolution of
the echo spectrometer is determined solely by the parameters of the
refocusing system, i.e., by the resolution of the hard x-ray
spectrograph \cite{Shvydko15}.  The parameters of the defocusing
system determine only the size of the secondary monochromatic source
on the sample $\Delta x_{\ind{1}}=|A_{\indrm{\mo}}|\Delta
x_{\ind{0}}$.

In the second step, we consider the particular optical designs of
x-ray echo spectrometers with a very high spectral resolution $\Delta
\etra \lesssim 0.1$~meV. For practical reasons, we will assume
that the secondary {\em monochromatic} source size is $\Delta
x_{\ind{1}}\simeq 5~\mu$m, and the focal length is
$f_{\ind{1}}\simeq 0.4$~m of the collimating element in the refocusing
system.  Then, Eq.~\eqref{resolution2} requires the ratio
${|\fcomm{\an}|} /{|\dcomm{\an}|}\simeq 125~\mu$rad/meV for the
dispersive element D$_{\indrm{\an}}$. Assuming the distance
$l_{\ind{3}}\simeq 2$~m from the focusing element to the sample in the
defocusing system, and $l_{\ind{1}} \gg \dcomm{\mo}^2
l_{\ind{2}}$, we estimate from Eq.~\eqref{refocus2} for
the required cumulative dispersing rate ${|\fcomm{\an}|} \simeq
25~\mu$rad/meV in the defocusing dispersive element. These are
relatively large values.  Typically, in a single Bragg reflection, a
maximum dispersion rate is $\dirate \simeq 6-8~\mu$rad/meV for photons
with energy $E\simeq 10$~keV \cite{SLK06,ShSS11}. As mentioned before,
multi-crystal arrangements can be used to enhance the dispersion rate
\cite{SSM13}.

Such large dispersion rates $\fcomm{}$, unfortunately, tend to
decrease the transmission bandwidths $\Delta E_{\ind{\cup}}$ of the
dispersing elements \cite{SSM13,Shvydko15}.  Achieving strong signals
in IXS experiments, however, requires $\Delta E_{\ind{\cup}} \gg
\Delta \etra$.  Therefore, optical designs of the dispersing
elements have to be found featuring both large $\fcomm{}$ and $\Delta
E_{\ind{\cup}}$.  Figures~\ref{fig003} and \ref{fig004} show
representative examples of multi-crystal CDDW-type inline dispersing
elements \cite{ShSS11,SSS13,SSS14} of the defocusing and refocusing
systems, respectively, with the required cumulative dispersion rates,
asymmetry factors, and with bandwidths $\Delta E_{\ind{\cup}}\simeq
5.5-9$~meV, i.e., $\Delta E_{\ind{\cup}}/\Delta \etra\simeq
55-90$, designed for use with 9.1-keV photons.
These examples are modifications of the dispersing elements 
designs presented in \cite{Shvydko15}. 

The CDDW optic in the ($\pi+$,$0-$,$0+$,$0-$) scattering configuration
is preferred for the defocusing dispersing element D$_{\indrm{\mo}}$
(Fig.~\ref{fig003}) as it provides the required dispersion rate
${|\fcomm{\mo}|}\simeq 25~\mu$rad/meV, significant transmission
bandwidth $\Delta E_{\ind{\cup}}\simeq 9$~meV, and is compact.  The
CDDW optic in the ($\pi+$,$\pi+$,$\pi-$,$0-$) configuration is better
suited for the refocusing dispersing element D$_{\indrm{\an}}$ (Fig.~\ref{fig004}). It provides the large ratio
  ${|\fcomm{\an}|} /{|\dcomm{\an}|}\simeq 120~\mu$rad/meV required for
  the high spectral resolution [see Eq.~\eqref{resolution2}] and
  substantial transmission bandwidth $\Delta E_{\ind{\cup}}\simeq
  5.5$~meV, although smaller than in the D$_{\indrm{\mo}}$-case.

  The total beam size on the sample is
    $\simeq G_{\indrm{\mo}}\Delta E_{\ind{\cup}}$. For the
    spectrometer exemplified here, it is estimated to
    be 
    $\simeq 275~\mu$m.
  Equation~\eqref{tolerances} together with
    Eqs.~\eqref{defocusing} and \eqref{refocusing} can be used to
    estimate tolerances on the admissible variations of the focal
    distances, sample displacement, sample surface imperfections,
    etc. As discussed in Appendix~\ref{echotollerances}  
the tolerances are
    in a millimeter range in the case of the 0.1-meV spectrometer.

Details on the optical designs and examples of the
dispersing elements designed for use with a lower photon energy of
4.57~keV and larger $\Delta E_{\ind{\cup}}\simeq 13-10$~meV, can be
found in   Appendix~\ref{cddwoptic}.
All these examples showcase the
applicability of echo spectroscopy in a wide spectral range, and its
feasibility both with synchrotron radiation and x-ray free electron
laser sources.

The point is that even higher spectral resolution $\Delta \etra
\lesssim 0.02$~meV can be achieved with x-ray echo spectrometers by
increasing the dispersion rates $\fcomm{}$ in the dispersing
elements. This, however, will result in their narrower transmission
bandwidths $\Delta E_{\ind{\cup}}$.  Still, an approximately constant
ratio $\Delta E_{\ind{\cup}}/\Delta \etra$ holds. Alternatively,
the spectral resolution can be improved by increasing the focal length
$f_{\ind{1}}$ in the refocusing system, see Eq.~\eqref{resolution2}.

The essential feature of the echo spectrometers is
that the signal strength, which is proportional to
  the product of the bandwidths of the photons on the sample and on
  the detector, is enhanced by $(\Delta E_{\ind{\cup}}/\Delta
\etra)^2 \simeq 10^3-10^4$ compared to what is possible with the
standard scanning-IXS-spectrometer approach.

\section{Conclusions}

In conclusion, x-ray echo spectroscopy, a counterpart of neutron
spin-echo, is introduced here to overcome limitations in spectral
resolution and weak signals of the traditional inelastic hard x-ray
scattering (IXS) probes. Operational principles, refocusing
conditions, and spectral resolutions of echo spectrometers are
substantiated by an analytical ray-transfer-matrix approach. A
principle optical scheme for a hard x-ray echo spectrometer is
proposed with multi-crystal arrangements as dispersing
elements. Concrete schemes are discussed with  5--13-meV transmission
bandwidths, a spectral resolution of 0.1-meV (extension to 0.02-meV is
realistic), and designed for use with 9.1-keV and 4.6-keV photons.  The
signal in echo spectrometers is enhanced by at least three orders of
magnitude compared to what is possible with the standard
scanning-IXS-spectrometer approach.

X-ray echo spectrometers require a combination of the CDDW dispersing
elements and focusing optics as major optical components. Such
components have been experimentally demonstrated recently
\cite{SSM13,SSS14}.  Implementation of x-ray echo spectrometers is,
therefore, realistic.

\section{Acknowledgments}
Stimulating discussions with D.-J.~Huang (NSRRC) are greatly
appreciated.  S.P.~Collins (DLS) is acknowledged for reading the
manuscript and for valuable suggestions.  Work at Argonne National
Laboratory was supported by the U.S. Department of Energy, Office of
Science, under Contract No. DE-AC02-06CH11357.


\appendix 
\newpage
~
\newpage

\section{Ray-transfer matrices} 
\label{raytransfer}

Ray-transfer matrices $\{A0G,CDF,001\}$ of the defocusing
$\hat{O}_{\indrm{D}}$ and refocusing $\hat{O}_{\indrm{R}}$ systems of
the x-ray echo spectrometers used in the paper are given in the last
two rows of Table~\ref{tab2}. They are equivalent to the derived in
Ref.~\cite{Shvydko15} ray-transfer matrices of x-ray focusing
monochromators and spectrographs. The matrices of the multi-element
systems $\hat{O}_{\indrm{D}}$ and $\hat{O}_{\indrm{R}}$ are obtained
by successive multiplication of the matrices of the constituent
optical elements, which are given in the upper rows of
Table~\ref{tab2}.  

\begin{figure}
\includegraphics[width=0.5\textwidth]{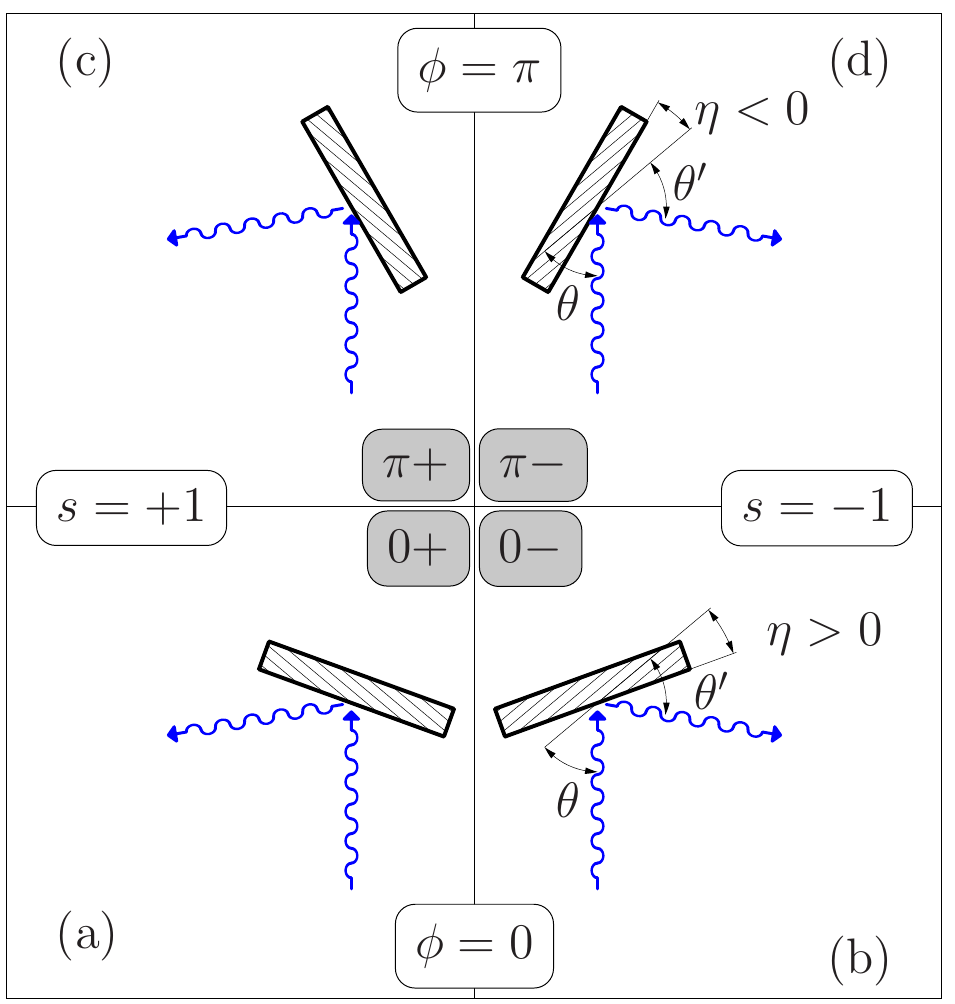}
\caption{Definitions of scattering geometries in Bragg diffraction from a crystal: (a) $0+$,  (b) $0-$,  (c) $\pi+$,  and (d) $\pi-$.}
\label{fig005}
\end{figure}

In the first three rows, 1--3, matrices are shown for the basic
optical elements, such as propagation in free space
$\hat{\fspace}(l)$, thin lens or focusing mirror $\hat{\thinlens}(f)$,
and Bragg reflection from a crystal $\hat{\crystal}(b,\sgn \dirate)$.
Scattering geometries in Bragg diffraction from crystals are defined
in Fig.~\ref{fig005}.  In the following rows of Table~\ref{tab2},
ray-transfer matrices are shown for arrangements composed of several
basic optical elements, such as successive multiple Bragg reflections
from crystals $\hat{\crystal}(\dcomm{n},\fcomm{n})$ and
$\hat{\crystalsp}(\dcomm{n},\fcomm{n},l)$, rows 4--5; and a focusing
system $\hat{\focus}(l_{\ind{2}},f,l_{\ind{1}})$, row 6.

The matrices of the defocusing $\hat{O}_{\indrm{D}}$ and refocusing
$\hat{O}_{\indrm{R}}$ systems presented in Table~\ref{tab2}, rows 7
and 8, respectively, are calculated using the multi-crystal matrix
$\hat{\crystal}(\dcomm{n},\fcomm{n})$ from row 4, assuming zero free
space between crystals in successive Bragg reflections.
Generalization to a more realistic case of nonzero distances between
the crystals requires the application of matrix
$\hat{\crystalsp}(\dcomm{n},\fcomm{n},l)$ from row 5.

We refer to Ref.~\cite{Shvydko15} for details on the derivation of
these matrices. Here, we provide only the final results, notations,
and definitions.

\begin{table*}
\centering
\begin{tabular}{|l|l|l|l|}
  \hline 
& & & \\
  Optical system & Matrix notation & Ray-transfer matrix  & Definitions and remarks\\[-5pt]    
& & & \\
  \hline  \hline
& & & \\[-3.8mm]  
\parbox[c]{0.25\textwidth}{Free space \cite{KL66,HoWe05,Siegman}\\  \includegraphics[width=0.25\textwidth]{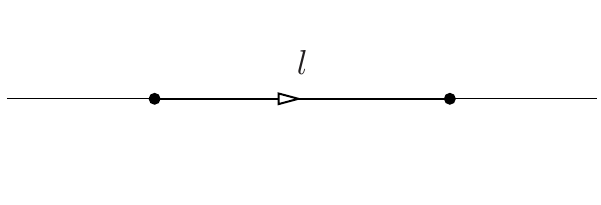}\\[-8mm] \hspace*{-40mm} (1)\\[3mm] }       &  $\hat{\fspace}(l)$ & $\left( \begin{array}{ccc} 1 & l  & 0 \\ 0 & 1 & 0 \\ 0 & 0 & 1  \end{array} \right)$      & \parbox[c]{0.21\textwidth}{$l$ -- distance}     \\[-0.5mm] 
\hline 
& & & \\[-3.8mm]  
\parbox[c]{0.25\textwidth}{Thin lens \cite{KL66,HoWe05,Siegman}\\ \includegraphics[width=0.25\textwidth]{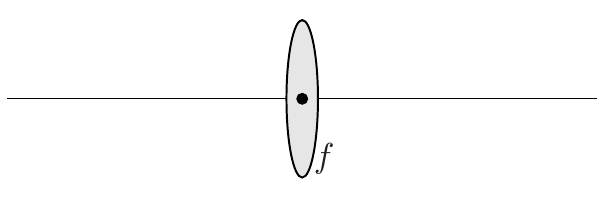}\\[-8mm] \hspace*{-40mm} (2)\\[3mm] }       &  $\hat{\thinlens}(f)$ & $\left( \begin{array}{ccc} 1 & 0 & 0\\ -\frac{1}{f} & 1 & 0  \\ 0 & 0 & 1 \end{array} \right)$      &  \parbox[c]{0.21\textwidth}{$f$ -- focal length}      \\ [-0.5mm]  
\hline 
& & & \\[-4.2mm]  
\parbox[c]{0.25\textwidth}{Bragg reflection from a crystal\\ \cite{MK80-1,MK80-2}\\ \includegraphics[width=0.25\textwidth]{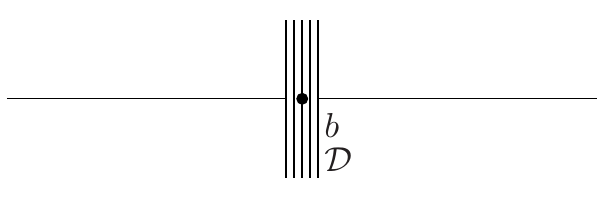} \\[-8mm] \hspace*{-40mm} (3)\\[3mm] }  &  $\hat{\crystal}(b,\sgn \dirate)$ & $\left( \begin{array}{ccc} {1}/{b} & 0 & 0  \\0  & b & \sgn\dirate \\ 0 & 0 & 1 \end{array} \right)$  &  \parbox[c]{0.21\textwidth}{$b=-\frac{\sin(\theta+\eta)}{\sin(\theta-\eta)}$\\ asymmetry factor;\\ $\dirate = -(1/E)(1+b)\tan\theta $\\ angular dispersion rate.
}     \\[+0.5mm] 
\hline 
& & & \\[-3.8mm]  
\parbox[c]{0.25\textwidth}{Successive Bragg reflections \cite{Shvydko15}\\ $\hat{\crystal}(\!b_{\ind{n}},\sgn_{\ind{n}}\dirate_{\ind{n}}\!)\dotsb \hat{\crystal}(\!b_{\ind{1}},\sgn_{\ind{1}}\dirate_{\ind{1}}\!)$ \\  \includegraphics[width=0.25\textwidth]{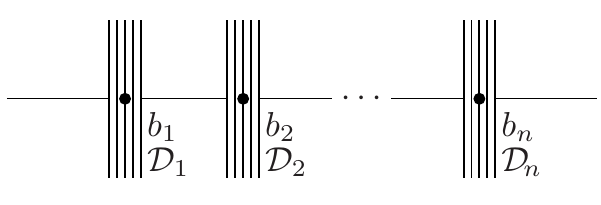} \\[-8mm] \hspace*{-40mm} (4)\\[3mm] }  &  $\hat{\crystal}(\dcomm{n},\fcomm{n})$ & $\left( \begin{array}{ccc} \acomm{n} & 0 & 0  \\ 0  & \dcomm{n}  & \fcomm{n} \\ 0 & 0 & 1 \end{array} \right)$  &  \parbox[c]{0.21\textwidth}{$\dcomm{n}=b_{\ind{1}}b_{\ind{2}}b_{\ind{3}} \dotsc b_{\ind{n}}$ \\ $\fcomm{n}=b_{\ind{n}}\fcomm{n-1} + \sgn_{\ind{n}}\dirate_{\ind{n}}$\\ $\sgn_i=\pm 1$, $i=1,2,...,n$ }     \\[-0.5mm] 
\hline 
& & & \\[-3.8mm]  
\parbox[c]{0.25\textwidth}{Successive Bragg reflections with space between crystals \cite{Shvydko15}\\ $\hat{\crystal}(\!b_{\ind{n}},\sgn_{\ind{n}}\dirate_{\ind{n}}\!)\dotsb \hat{\fspace}(\!l_{\ind{12}}\!)\hat{\crystal}(\!b_{\ind{1}},\sgn_{\ind{1}}\dirate_{\ind{1}}\!)$ \\  \includegraphics[width=0.25\textwidth]{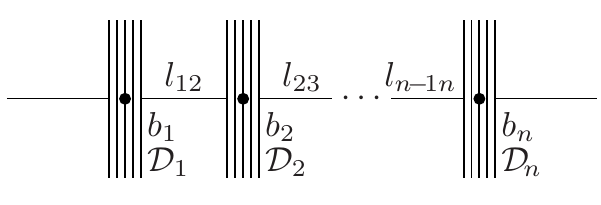}\\[-8mm] \hspace*{-40mm} (5)\\[3mm] }  &  $\hat{\crystalsp}(\dcomm{n},\fcomm{n},l)$ & $\left( \begin{array}{ccc} \acomm{n} & \bcomm{n} & \gcomm{n}  \\ 0  & \dcomm{n}  & \fcomm{n} \\ 0 & 0 & 1 \end{array} \right)$  &  \parbox[c]{0.21\textwidth}{$\bcomm{n}\!=\!\frac{\bcomm{n-1}+\dcomm{n-1}l_{\ind{n-1 n}}}{b_{\ind{n}}}$\\[1mm]
$\gcomm{n}\!=\!
\frac{\gcomm{n-1}+\fcomm{n-1}l_{\ind{n-1 n}}}{b_{\ind{n}}}$\\[1mm]  $\bcomm{1}\!=\!0, \hspace{0.5cm} \gcomm{1}\!=\!0 $}     \\[-0.5mm] 
\hline 
& & & \\[-3.8mm]
\parbox[c]{0.25\textwidth}{Focusing system\\ $\hat{\fspace}(l_{\ind{2}})\hat{\thinlens}(f)\hat{\fspace}(l_{\ind{1}})$ \\ \includegraphics[width=0.25\textwidth]{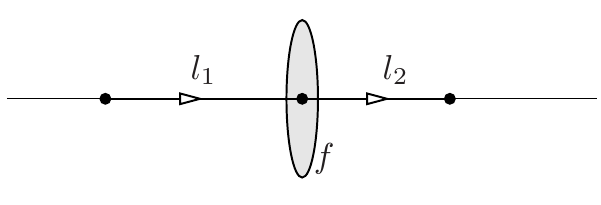} \\[-8mm] \hspace*{-40mm} (6)\\[3mm] }  &  $\hat{\focus}(l_{\ind{2}},f,l_{\ind{1}})$ & $\left( \begin{array}{ccc} 1-\frac{l_{\ind{2}}}{f} & B_{\indrm{F}} & 0  \\ -\frac{1}{f}  & 1-\frac{l_{\ind{1}}}{f} & 0\\ 0 & 0 & 1  \end{array} \right)$  &  \parbox[c]{0.21\textwidth}{$B_{\indrm{F}}=l_{\ind{1}}l_{\ind{2}}\left(\frac{1}{l_{\ind{1}}}+\frac{1}{l_{\ind{2}}}-\frac{1}{f}\right)$ }     \\[-0.5mm] 
\hline 
\hline 
& & & \\[-3.8mm]
\parbox[c]{0.25\textwidth}{Defocusing system $\hat{O}_{\indrm{D}}$ \cite{Shvydko15}\\ $\hat{\focus}(l_{\ind{3}},f,l_{\ind{2}}) \hat{\crystal}(\dcomm{n}\!,\fcomm{n}\!)\hat{\fspace}(l_{\ind{1}})$ \\ \includegraphics[width=0.25\textwidth]{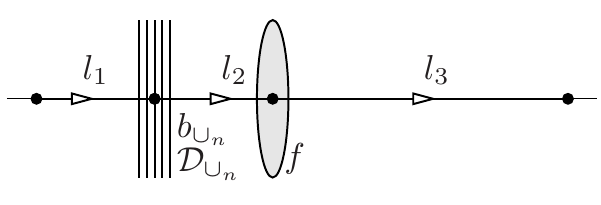} \\[-8mm] \hspace*{-40mm} (7)\\[3mm] }  &  $\hat{O}_{\indrm{D}}$ & $\left(\!\! \begin{array}{ccc} \frac{1}{\dcomm{n}\!}\!\left(\!1\!-\!\frac{l_{\ind{3}}}{f}\!\right) & 0  &  X \fcomm{n}  \\ 
-\frac{1}{f\dcomm{n}}  & \dcomm{n}\!\!\left(\!1\!-\!\frac{l_{12}}{f}\!\right) & \left(\!1\!-\!\frac{l_{\ind{2}}}{f}\!\right)\!\fcomm{n} \\ 
0 & 0 & 1  \end{array}\!\! \right)
 $  &  \parbox[c]{0.21\textwidth}{$\frac{1}{l_{\ind{12}}}\!+\!\frac{1}{l_{\ind{3}}}\!=\!\frac{1}{f}$ \\[1mm]
$l_{\ind{12}} =l_{\ind{1}} /\dcomm{n}^2 + l_{\ind{2}}  $ \\[1mm] $X\!=\!l_{\ind{3}} l_{\ind{1}}\!/(\!\dcomm{n}^2\! l_{\ind{12}}\!) $  }     \\[-0.5mm] 
\hline 
& & & \\[-3.8mm]
\parbox[c]{0.25\textwidth}{Refocusing system $\hat{O}_{\indrm{R}}$ \cite{Shvydko15}\\ $\hat{\focus}(\!f_{\ind{2}}\!,\!f_{\ind{2}}\!,\!l_{\ind{2}}\!) \hat{\crystal}(\!\dcomm{n}\!,\!\fcomm{n}\!) \hat{\focus}(\!l_{\ind{1}}\!,\!f_{\ind{1}}\!,\!f_{\ind{1}}\!) $\\ \includegraphics[width=0.25\textwidth]{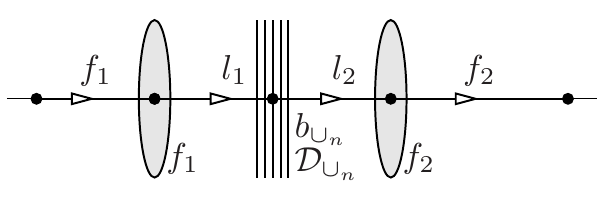} \\[-8mm] \hspace*{-40mm} (8)\\[3mm] }   &  $\hat{O}_{\indrm{R}}$ & $ 
\left(\!\! \begin{array}{ccc} -\frac{\dcomm{n}f_{\ind{2}}\!}{f_{\ind{1}}\!}  & 0 &  f_{\ind{2}} \fcomm{n}  \\ \frac{(\!l_{\ind{1}}\!-\!f_{\ind{1}}\!)\!+\!(\!l_{\ind{2}}\!-\!f_{\ind{2}}\!)\dcomm{n}^2\!}{\!\dcomm{n}f_{\ind{1}}f_{\ind{2}}\!}  & -\frac{f_{\ind{1}}}{\dcomm{n}\!f_{\ind{2}}} & \left(\!1\!-\!\frac{l_{\ind{2}}}{f_{\ind{2}}}\!\right)\!\fcomm{n}\! \\ 0 & 0 & 1 \end{array}\!\! \right)
 $  &  \parbox[c]{0.21\textwidth}{ }     \\ 
  \hline  
\end{tabular}
\caption{Ray-transfer matrices $\{ABG,CDF,001\}$ of optical systems
  used in the paper. The matrices are shown starting with basic ones
  in rows 1--3. Matrices of combined systems are given in rows
  4--6. The ray transfer matrices of the defocusing
  $\hat{O}_{\indrm{D}}$ and refocusing $\hat{O}_{\indrm{R}}$ systems
  of x-ray echo spectrometers are presented in rows 7--8. Definition
  of the glancing angle of incidence $\theta$ to the reflecting
  crystal atomic planes, the asymmetry angle $\eta$, and the
  deflection sign $s$ in Bragg diffraction from a crystal, used for
  the Bragg reflection ray-transfer matrix in row 3, are given in
  Fig.~\ref{fig005}. See Ref.~\cite{Shvydko15} for more details.  }
\label{tab2}
\end{table*}


\section{CDDW optic as dispersing element}  
\label{cddwoptic}

\begin{figure}
\includegraphics[width=0.5\textwidth]{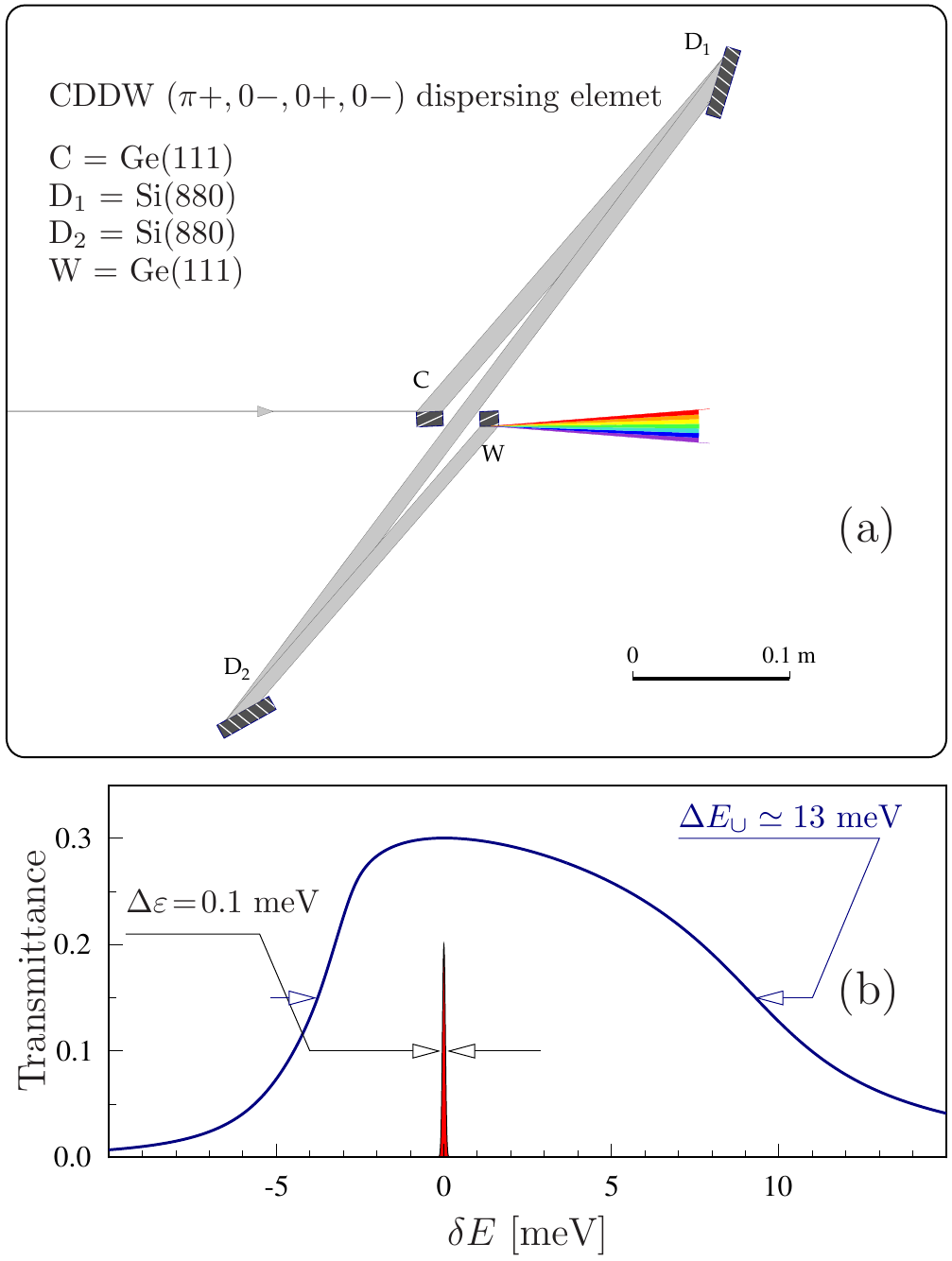}
\caption{Dispersing element D$_{\indrm{\mo}}$ (a) of the defocusing
  system $\hat{O}_{\indrm{\mo}}$ (Fig.~2) and its spectral
  transmittance function (b).  D$_{\indrm{\mo}}$ is an example of an
  in-line four-crystal CDDW-type dispersing optic, comprised of
  collimating (C), dispersing (D$_{\ind{1}}$, D$_{\ind{2}}$), and
  wavelength-selecting (W) crystals in a ($\pi+$,$0-$,$0+$,$0-$)
  scattering configuration. With the crystal parameters provided in
  Table~\ref{tab14}, the dispersing element D$_{\indrm{D}}$ features a
  spectral transmission function with a $\Delta E_{\ind{\cup}}=13$~meV
  bandwidth (b), a cumulative angular dispersion rate $\fcomm{\mo} =
  -42 ~\mu$rad/meV, and a cumulative asymmetry factor $\dcomm{\mo} =
  1.4$.  The sharp line in (b) presents the 0.1-meV design spectral
  resolution of the x-ray echo spectrometer. $E=4.5686$~keV. }
\label{fig00399}
\end{figure}

\begin{figure}
\includegraphics[width=0.5\textwidth]{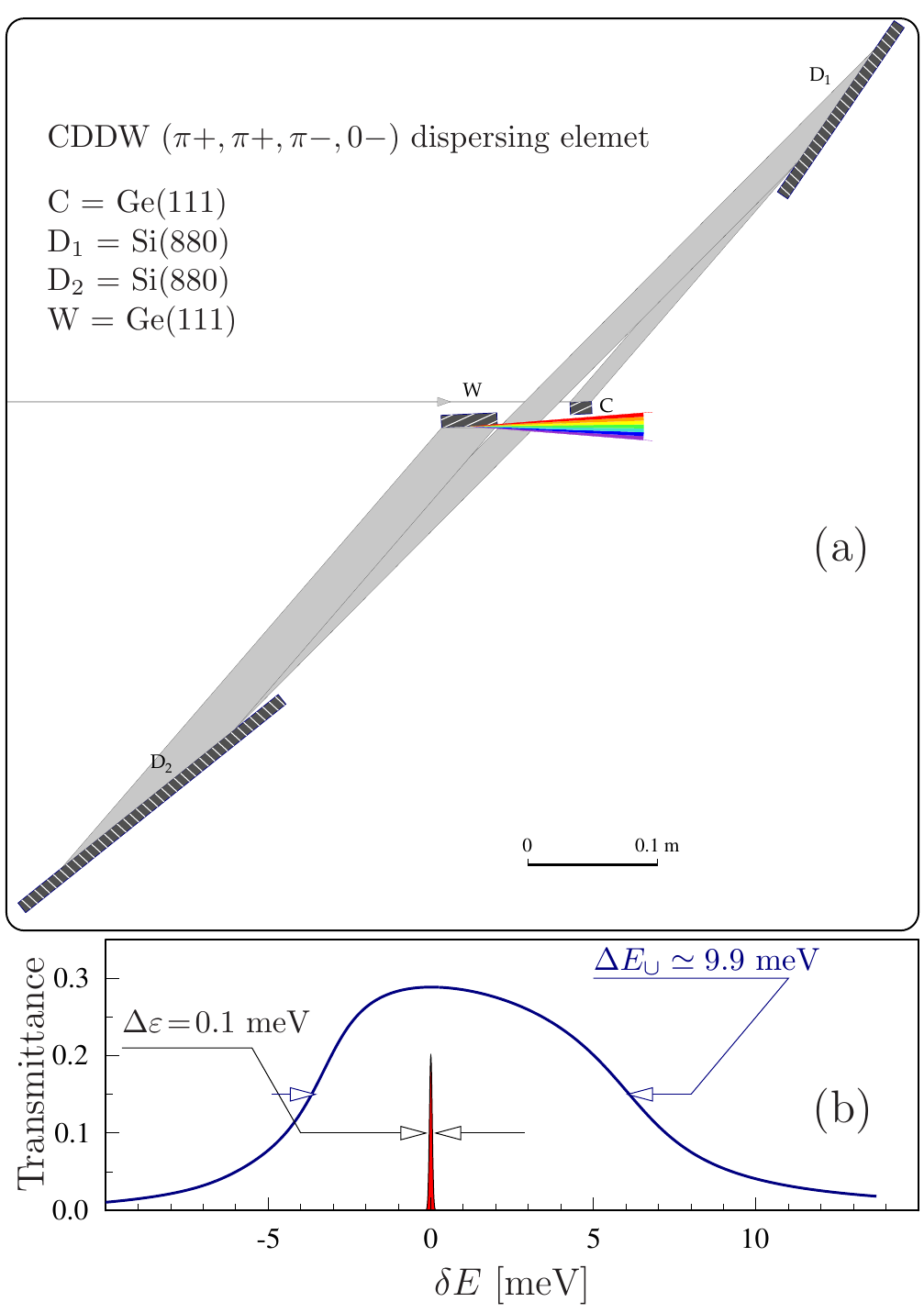}
\caption{Dispersing element D$_{\indrm{\an}}$ (a) of the refocusing
  system $\hat{O}_{\indrm{\an}}$ (Fig.~2) and its spectral
  transmittance function (b).  
Similar to D$_{\indrm{\mo}}$ in
  Fig.~\ref{fig00399}, D$_{\indrm{\an}}$ is an example of the in-line
  four-crystal CDDW-type dispersing optic, but, in a
  ($\pi+$,$\pi+$,$\pi-$,$0-$) scattering configuration. With the
  crystal parameters provided
  in Table~\ref{tab14}, the dispersing element D$_{\indrm{D}}$ features a
  spectral transmission function with a $\Delta
  E_{\ind{\cup}}=9.9$~meV bandwidth (b), a cumulative angular
  dispersion rate $\fcomm{\mo} = -64 ~\mu$rad/meV, a cumulative
  asymmetry factor $\dcomm{\mo} = 0.39$, and $\fcomm{\mo}/\dcomm{\mo}
  = -161 ~\mu$rad/meV.  The sharp line in (b) presents the 0.1-meV
  design spectral resolution of the x-ray echo
  spectrometer. $E=4.5686$~keV.  }
\label{fig0049}  
\end{figure}

In-line four-crystal CDDW-type dispersing optics
\cite{ShSS11,SSS13,SSS14} are proposed in the paper for use as
dispersing elements D$_{\indrm{D}}$, D$_{\indrm{R}}$ of the defocusing
$\hat{O}_{\indrm{D}}$ and refocusing systems $\hat{O}_{\indrm{\an}}$
of the echo spectrometer, respectively.  The in-line four-crystal
CDDW-type dispersing optic (see schematics in Figs.~3-4 and
Figs.~\ref{fig00399}-\ref{fig0049}), comprises collimating (C),
dispersing (D$_{\ind{1}}$, D$_{\ind{2}}$), and wavelength-selecting
(W) crystals, which can be arranged in different scattering
configurations. In a general case, the scattering configuration is
defined as
$(\phi_{\ind{1}}s_{\ind{1}},\phi_{\ind{2}}s_{\ind{2}},\phi_{\ind{3}}s_{\ind{3}},\phi_{\ind{4}}s_{\ind{4}})$.
Here, for each crystal $n=1,2,3,4$ (1=C, 2=D$_{\ind{1}}$,
3=D$_{\ind{2}}$, 4=W) the value $\phi_{\ind{n}}=0$ corresponds to
the grazing reflection, see Fig.~\ref{fig005}(a)-(b); while
$\phi_{\ind{n}}=\pi$ corresponds to the grazing incidence, see
Fig.~\ref{fig005}(c)-(d). The sign $s_{\ind{n}}=+1$ corresponds to
a reflection in the counterclockwise direction, see Figs.~\ref{fig005}(a),(c);
while $s_{\ind{n}}=-1$ means the clockwise direction, see
Figs.~\ref{fig005}(b),(d).  Those scattering geometries have been
selected for use as dispersing elements of the x-ray echo spectrometer
in the paper, which feature the largest cumulative dispersion rates
$\fcomm{n}$.

The cumulative dispersion rate $\fcomm{4}$ in a four-crystal system is
given in a general case by
\begin{equation}
\fcomm{4} =
b_{\ind{4}}b_{\ind{3}}b_{\ind{2}}\sgn_{\ind{1}}\dirate_{\ind{1}}+b_{\ind{4}}b_{\ind{3}}\sgn_{\ind{2}}\dirate_{\ind{2}}+b_{\ind{4}}\sgn_{\ind{3}}\dirate_{\ind{3}}+\sgn_{\ind{4}}\dirate_{\ind{4}},
\end{equation}
with the asymmetry parameters $b$ and dispersion rates $\dirate$
defined in Table~\ref{tab2}, rows (3) and (4).  For the CDDW-type
dispersing elements considered in this paper the dispersion rate of
the D-crystals ($n=2,3$) is much larger than those of the C- and
W-crystals ($n=1,4$), see Tables~\ref{tab1} and \ref{tab14}.  In this
case, the cumulative dispersion rate can be approximated by $\fcomm{4}
\simeq
b_{\ind{4}}b_{\ind{3}}\sgn_{\ind{2}}\dirate_{\ind{2}}+b_{\ind{4}}\sgn_{\ind{3}}\dirate_{\ind{3}}=b_{\ind{4}}(b_{\ind{3}}\sgn_{\ind{2}}\dirate_{\ind{2}}+\sgn_{\ind{3}}\dirate_{\ind{3}})$.
The largest dispersion rates can be achieved in systems, in which the
product is
$\sgn_{\ind{2}}\sgn_{\ind{3}}\dirate_{\ind{2}}\dirate_{\ind{3}}<0$.

There are four high-dispersion-rate CDDW configurations featuring
$\dirate_{\ind{2}}\dirate_{\ind{3}}>0$ and
$\sgn_{\ind{2}}\sgn_{\ind{3}}<0$: $(\pi+,\pi-,\pi+,0+)$;
$(\pi+,\pi+,\pi-,0-)$; $(\pi+,0-,0+,0-)$; and $(\pi+,0+,0-,0-)$. These
configurations are especially interesting because of the incident and
transmitted x-rays being parallel (in-line scheme).

There are four other high-dispersion-rate CDDW configurations featuring
$\dirate_{\ind{2}}\dirate_{\ind{3}}<0$ and
$\sgn_{\ind{2}}\sgn_{\ind{3}}>0$: $(\pi+,\pi-,0-,0-)$;
$(\pi+,\pi+,0+,0-)$; $(\pi+,0-,\pi-,0-)$; and  $(\pi+,0+,\pi+,0-)$.
However, they are not in-line. The angle between the incident and
reflected beams is $4(\theta_{\mathrm D}-\pi/2)$.


In the present paper, we have chosen the in-line high-dispersion-rate
CDDW optic in the ($\pi+$,$0-$,$0+$,$0-$) configuration, as a
dispersing element D$_{\indrm{\mo}}$ of the defocusing system
$\hat{O}_{\indrm{\mo}}$, see Figs.~3 and \ref{fig00399}.  The in-line
high-dispersion-rate CDDW optic in the ($\pi+$,$\pi+$,$\pi-$,$0-$)
configuration, was chosen as a dispersing element D$_{\indrm{\an}}$ of
the refocusing system $\hat{O}_{\indrm{\an}}$, see Figs.~4 and
\ref{fig0049}.

Tables~\ref{tab1} and \ref{tab14} present crystal parameters and
cumulative parameters of the dispersing elements designed for
operations with x-rays with photon energies of 9.131385~keV and
4.5686~keV, respectively.


\begin{table}
\centering
\begin{tabular}{|l|lllllll|}
  \hline   \hline 
crystal   & $\vc{H}_{\indrm{\elmt}}$ &$\eta_{\indrm{\elmt}} $ &$\theta_{\indrm{\elmt}} $  & $\dei{\elmt} $ &  $\dai{\elmt}$  & $b_{\indrm{\elmt}}$ & $s_{\indrm{\elmt}}\dirate_{\indrm{\elmt}} $ \\[-5pt]    
element (\elmt)  & &  &  &   &  &   &    \\[0pt]    
[material]          & $(hkl)$ & deg & deg  & meV  &  $\mu$rad   & & $\frac{\mu {\mathrm {rad}}}{\mathrm {meV}}$ \\[5pt]    
\hline  \hline  
\multicolumn{8}{|l|}{D$_{\indrm{\mo}}$: CDDW ($\pi+$,$0-$,$0+$,$0-$), Fig.~3}\\  
\cline{1-1}
1~~C~~[Ge]& (1~1~1) &  -10.0  &  12.0  &  3013 & 71    & -0.09  & -0.02 \\[-0.0pt]
2~~D$_{\ind{1}}$~[Si] & (8~0~0) &  77.5  &  89  &  27 &  341    & -1.17  & -1.07 \\[-0.0pt]
3~~D$_{\ind{2}}$~[Si] & (8~0~0) &  77.5  &  89   & 27 &  341   & -1.17  & +1.07 \\[-0.0pt]
4~~W~~[Ge] & (1~1~1) &  10.0  &  12.0   & 3013 &  71   & -10.8  & -0.22 \\[-0.0pt]
  \hline  
\multicolumn{4}{|l}{}  & $\Delta E_{\ind{\cup}}$  & $\Delta \theta_{\ind{\cup}}^{\prime}$  & $\dcomm{ }$ & $\fcomm{ }$ \\
\multicolumn{4}{|l}{}  & meV  & $\mu$rad  &  & $\frac{\mu {\mathrm {rad}}}{\mathrm {meV}}$ \\[5pt]    
  \hline  
\multicolumn{4}{|l}{Cumulative values}  & 8.8  & -218  & 1.38 & -25.0 \\
\hline  \hline  
\multicolumn{8}{|l|}{D$_{\indrm{\an}}$: CDDW ($\pi+$,$\pi+$,$\pi-$,$0-$), Fig.~4}\\  
\cline{1-1}
1~~C~~[Ge]& (1~1~1) &  -10.0  &  12.0  &  3013 & 71    & -0.09  & -0.02 \\[-0.0pt]
2~~D$_{\ind{1}}$~[Si] & (8~0~0) &  -86  &  89  &  27 &  341    & -0.6  & -2.50 \\[-0.0pt]
3~~D$_{\ind{2}}$~[Si] & (8~0~0) &  -86  &  89   & 27 &  341   & -0.6  & +2.50 \\[-0.0pt]
4~~W~~[Ge] & (1~1~1) &  10.0  &  12.0   & 3013 &  71   & -10.8  & -0.22 \\[-0.0pt]
  \hline  
\multicolumn{4}{|l}{Cumulative values}  & 5.5  & -237  & 0.36 & -43.5 \\
  \hline  
  \hline  
\end{tabular}
\caption{Examples of the in-line four-crystal CDDW optics as dispersing elements 
  (``diffraction gratings'') 
  D$_{\indrm{D}}$, D$_{\indrm{R}}$ of the defocusing
  $\hat{O}_{\indrm{D}}$ and refocusing $\hat{O}_{\indrm{\an}}$ systems 
  of the echo spectrometer, respectively.
  For each optic, the table presents crystal elements (\elmt=C,D$_{\ind{1}}$,D$_{\ind{2}}$,W) and their Bragg reflection parameters:
  $(hkl)$, Miller indices of the Bragg diffraction vector $\vc{H}_{\indrm{\elmt}}$;  
  $\eta_{\indrm{\elmt}}$, asymmetry angle; $\theta_{\indrm{\elmt}}$, glancing angle of incidence; $\deis{\elmt}$, $\dais{\elmt}$
  are Bragg's reflection intrinsic spectral width and angular acceptance in symmetric scattering geometry, respectively; 
  $b_{\indrm{\elmt}}$, asymmetry factor; and  $s_{\indrm{\elmt}} \dirate_{\indrm{\elmt}} $, angular dispersion rate with deflection sign.  For each optic, the table also shows  the 
  spectral window of imaging $\Delta E_{\ind{\cup}}$ as derived from the dynamical  theory calculations, the angular spread of the dispersion fan 
  $\Delta \theta_{\ind{\cup}}^{\prime}=\fcomm{ } \Delta E_{\ind{\cup}}$, 
  and the cumulative values of the asymmetry parameter $\dcomm{ }$  and the dispersion rate $\fcomm{ }$.
  X-ray photon energy is $E=9.13185$~keV in both cases.
}
\label{tab1}
\end{table}

\begin{table}
\centering
\begin{tabular}{|l|lllllll|}
  \hline   \hline 
crystal   & $\vc{H}_{\indrm{\elmt}}$ &$\eta_{\indrm{\elmt}} $ &$\theta_{\indrm{\elmt}} $  & $\dei{\elmt} $ &  $\dai{\elmt}$  & $b_{\indrm{\elmt}}$ & $s_{\indrm{\elmt}}\dirate_{\indrm{\elmt}} $ \\[-5pt]    
element (\elmt)  & &  &  &   &  &   &    \\[0pt]    
[material]          & $(hkl)$ & deg & deg  & meV  &  $\mu$rad   & & $\frac{\mu {\mathrm {rad}}}{\mathrm {meV}}$ \\[5pt]    
\hline  \hline  
\multicolumn{8}{|l|}{D$_{\indrm{\mo}}$: CDDW ($\pi+$,$0-$,$0+$,$0-$), Fig.~\ref{fig00399}}\\  
\cline{1-1}
1~~C~~[Ge]& (1~1~1) &  -22.0  &  24.60  &  1542 & 154    & -0.06  & -0.09 \\[-0.0pt]
2~~D$_{\ind{1}}$~[Si] & (4~0~0) &  68  &  88  &  110 &  691    & -1.19  & -1.18 \\[-0.0pt]
3~~D$_{\ind{2}}$~[Si] & (4~0~0) &  68  &  88   & 110 &  691   & -1.19  & +1.18 \\[-0.0pt]
4~~W~~[Ge] & (1~1~1) &  22.0  &  24.55   & 1542 &  154   & -16.4  & -1.53 \\[-0.0pt]
  \hline  
\multicolumn{4}{|l}{Cumulative values}  & 13.0  & -547  & 1.44 & -41.9 \\
  \hline  
  \hline  
\multicolumn{8}{|l|}{D$_{\indrm{\an}}$: CDDW  ($\pi+$,$\pi+$,$\pi-$,$0-$), Fig.~\ref{fig0049}} \\  
\cline{1-1}
1~~C~~[Ge]& (1~1~1) &  -22.0  &  24.60  &  1542 & 154    & -0.06  & -0.09 \\[-0.0pt]
2~~D$_{\ind{1}}$~[Si] & (4~0~0) &  -81.5  &  88  &  110 &  691    & -0.62  & -2.37 \\[-0.0pt]
3~~D$_{\ind{2}}$~[Si] & (4~0~0) &  -81.5  &  88   & 110 &  691   & -0.62  & +2.37 \\[-0.0pt]
4~~W~~[Ge] & (1~1~1) &  22.0  &  24.55   & 1542 &  154   & -16.4  & -1.53 \\[-0.0pt]
  \hline  
\multicolumn{4}{|l}{Cumulative values}  & 9.9  & -638  & 0.39 & -64.0 \\
  \hline  
  \hline  
\end{tabular}
\caption{Same as Table~\ref{tab1}, however with  the multi-crystal  CDDW dispersing elements designed for 
  an x-ray photon energy  $E=4.5686$~keV.
}
\label{tab14}
\end{table}

\newpage

\section{Focusing and collimating optics}  

Focusing and collimating optic elements are another key components of
the x-ray echo spectrometers. There are no principal preferences of
using either curved mirrors or compound-refractive lenses (CRL) for
this purpose. However, in practical terms, mirrors maybe a preferable
choise ensuring higher efficiency, because photoabsorption of 9.1-keV
and especially of 4.5-keV photons is substantial in the CRLs
\cite{SKSL,LST99}.

The focusing element $f$ in the defocusing dispersing system
$\hat{O}_{\indrm{\mo}}$ (see Fig.~2) can be a standard K-B mirror
system, ensuring a $\Delta x_{\ind{1}}\simeq 5~\mu$m vertical spot
size in the echo-spectrometer example considered in the present paper.
Tight focusing in the horizontal plane is also advantageous to
minitage the negative effect on the spectral resolution of a
"projected" scattering source size with increasing scattering angle.

The collimating element $f_{\ind{1}}$ in the refocusing dispersing
system $\hat{O}_{\indrm{\an}}$ collects photons in a large solid angle
$\Omega_{\indrm{h}}\times\Omega_{\indrm{v}}$, with
$\Omega_{\indrm{h}}\simeq 10$~mrad, $\Omega_{\indrm{h}}\simeq
1-10$~mrad (depending on the required momentum transfer resolution,
and makes the x-ray beam parallel. Laterally graded multilayer Montel
mirrors recently proved to be useful exactly in this role
\cite{MSL13,SSS14}.

The focusing element $f_{\ind{2}}$ in the refocusing dispersing system
$\hat{O}_{\indrm{\an}}$ focuses x-rays in the vertical dispersion
plane onto the detector. Because the vertical beamsize after the
dispersing element D$_{\indrm{\an}}$ is increased by $1/\dcomm{\mo}$
(an inverse of the cumulative asymmetry factor $\dcomm{\mo}$) the
$f_{\ind{2}}$ element has to have a relatively large vertical
geometrical aperture $\simeq 2-5$~mm (depending on the required
spectral resolution). One-dimensional parabolic mirrors should be able
to deal effectively with this problem.

\section{Echo spectrometer tolerances}
\label{echotollerances}

Tolerances on the echo spectrometer parameters can be calculated from
Eq.~(7) in a general case.  The equation can be rewritten as
\begin{equation}
|G_{\indrm{\mo}}+G_{\indrm{\an}}/A_{\indrm{\an}}| \Delta E_{\cup}\ll \Delta x_{\ind{1}}
\label{tollerances2} 
\end{equation}
using Eq.~(3) and the relationship $\Delta
x_{\ind{2}}= |A_{\indrm{\an}} | \Delta x_{\ind{1}}$ from Eq.~(4),

In a particular case of the echo spectrometer, which optical scheme is
shown in Fig.~2, the tolerances on the spectrometer parameters can be
calculated from equation
\begin{equation}
\left| \fcomm{\mo}  \frac{l_{\ind{3}} l_{\ind{1}}}{\dcomm{\mo}^2 l_{\ind{12}}} - \frac{  \fcomm{\an} f_{\ind{1}}\!  }{ \dcomm{\an}\!} \right| \Delta E_{\cup}\ll \Delta x_{\ind{1}}
\label{tolerances2} 
\end{equation}
which is obtained by combining Eq.~\eqref{tolerances2} and Eqs.(8)-(9). 
If the dispersing element D$_{\indrm{\mo}}$ is placed from the source
at a large distance $l_{\ind{1}} \gg \dcomm{\mo}^2 l_{\ind{2}}$, in
this case, the tolerance equation simplifies to
\begin{equation}
\left| \fcomm{\mo} l_{\ind{3}}  - \frac{  \fcomm{\an} f_{\ind{1}}\!  }{ \dcomm{\an}\!} \right| \Delta E_{\cup}\ll \Delta x_{\ind{1}}.
\label{tolerances3} 
\end{equation}
As an example, we assume that the spectrometer parameters are
perfectly adjusted, except for the distance $l_{\ind{3}}$ from the
focusing mirror to the secondary source (to the sample). The
tolerance interval $\Delta l_{\ind{3}}$  in this case can be estimated
using Eq.~\eqref{tolerances3} as
\begin{equation}
\left|\Delta l_{\ind{3}} \right| \ll \frac{\Delta x_{\ind{1}}}{ |\fcomm{\mo}| \Delta E_{\cup} }.
\label{toleranceintervall3} 
\end{equation}
If the distance $f_{\ind{1}}$ from the secondary source (sample) to
the collimating mirror is not perfectly adjusted, the tolerance
interval $\Delta f_{\ind{1}}$ can be estimated in this case as
\begin{equation}
\left|\Delta  f_{\ind{1}} \right| \ll \frac{\Delta x_{\ind{1}} \dcomm{\an}}{ |\fcomm{\an}| \Delta E_{\cup} }.
\label{toleranceintervalf1} 
\end{equation}
With the parameters of the 0.1-meV-resolution echo spectrometer
provided in the paper ($\Delta x_{\ind{1}}=5~\mu$m; $\Delta E_{\cup}
=5.5$~meV; $|\fcomm{\mo}|=25~\mu$rad/meV;
$|\dcomm{\an}/\fcomm{\an}|=125~\mu$rad/meV), these tolerance intervals
are estimated to be $\left|\Delta l_{\ind{3}} \right| \ll 36$~mm, and
$\left|\Delta f_{\ind{1}} \right| \ll 7$~mm, respectively. These
numbers are not extremely demanding. 

Since the variations of $l_{\ind{3}}$ and $f_{\ind{1}}$ could be
related to sample position displacement and surface imperfections or
to the sample being installed at some angle to the incident beam, the
above estimated numbers also provide constraints on imperfections in
the sample shape in this particular case.

\section{Tuning the refocusing condition up} 
In practice, the refocusing condition given by Eq.~(10) can be
exactly satisfied by tuning the distance $l_{\ind{3}}$, between the
focusing element $f$ and the sample, see Fig.~2. Given that the
source-to-sample distance $l=l_{\ind{1}}+l_{\ind{2}}+l_{\ind{3}}$, as
well as the focal distance
$f=l_{\ind{12}}l_{\ind{3}}/(l_{\ind{12}}+l_{\ind{3}})$ are fixed, the
distances $l_{\ind{1}}$ and $l_{\ind{2}}$ also have to be corrected by
the positioning of the dispersing system D$_{\indrm{\mo}}$,
appropriately. The distances $l_{\ind{1}}$ and $l_{\ind{2}}$ are
defined from the above mentioned constraints, by solving the equations
\begin{align}
l_{\ind{1}}+l_{\ind{2}}& = l-l_{\ind{3}}\\
\frac{l_{\ind{1}}}{\dcomm{\mo}^2}+l_{\ind{2}}& = \frac{f l_{\ind{3}}}{l_{\ind{3}}-f}.
\end{align}

\section{Spectral window of imaging} 
The spectral window of the imaging of the echo spectrometer is defined
by the bandwidths and their relative shifts of the defocusing and
refocusing systems. The shape of the window of imaging can be measured
by measuring the elastically scattered signal and scanning one
bandwidth against another. The window of imaging can be shifted by
shifting one of the bandwidths against another.

\end{document}